\documentclass[aps, prb, twocolumn]{revtex4}
\usepackage{ams}
\usepackage{graphicx}
\usepackage{color}

\begin{document}

\newcommand{\CdMnTe}[0]{\mbox{CdMnTe}}
\newcommand{\GaMnAs}[0]{\mbox{GaMnAs}}

\title{Local field of magnetic islands: role of their shape}
\author{Pawe{\l} Redli\'{n}ski}
\email{pawel.redlinski.1@nd.edu} \affiliation{Department of Physics,
University of Notre Dame, Notre Dame, Indiana
46556}

\begin{abstract}
I analyze in details distribution of local magnetic field induced by
micro- and nano-magnets. I consider three kinds of elongated
magnetic islands: ellipse-, diamond- and rectangular shaped islands
which were magnetized uniformly along long axis. This report
concentrates on the role of their shape upon distribution of the
field. Calculations show that unlike rectangular-shaped magnet,
ellipse-shaped and diamond-shaped ones produce much more localized
field in proximity of its magnetic poles. Additionally in the case
of ellipse-shaped islands the magnitude of induced field is large.
This two facts favor arrays of ellipse-shaped magnetic islands to
build zero-dimensional spin traps in a hybrid based on
Ferromagnet/Semiconductor structure.
\end{abstract}


\maketitle


\section{Introduction}
Basic research in the physical sciences, especially in condensed matter physics, can
result in important developments in applied physics and engineering. Recently
scientists draw theirs attention to the physics of hybrid structures. Both
(Ferromagnet)/(Magnetic Semiconductor)\cite{Crowell,Tanaka} and
(Superconductor)/(Magnetic Semiconductor),\cite{Tatiana,Castellana}  are promising
hybrids that join magnetic and electronic properties in one unit. Using Magnetic
Semiconductor or more precisely Diluted Magnetic Semiconductor (DMS), e.g., \CdMnTe{},
\GaMnAs{} instead of a Conventional alloy (CdTe, GaAs) amplifies some physical
properties which are extremely small in a (Ferromagnet or Superconductor)/(Conventional
Semiconductor) hybrid\cite{Michel}. The effect that lies the foundation of such
amplification is called Giant Zeeman Effect\cite{Furdyna1} and is due to the sp-d
exchange interaction between delocalized (s, p) and localized (d) carries. Let us
concentrate on the (Ferromagnet)/(DMS) hybrid. When DMS is buried below array of
magnetic islands, local field produced by each island is penetrating DMS structure. DMS
acts itself as an amplifier and multiplies this field by huge factor (typically by
hundred times but in sub-Kelvin temperatures even by thousand times). In this way
quasi-particles fill large, effective "potential" proportional to this local field and
which acts as a trap for them. Depending on the local distribution of the magnetic
field the effective potential can reveal one-dimensional\cite{Redlinski1} or
zero-dimensional\cite{Redlinski2} character. In this report I analyze the role of shape
of the islands\cite{Lebecki} on the distribution of induced field so on possibility of
localization of quasi-particles.

Micro- and nano-magnets of different kind of shapes have already
been investigated experimentally: diamond-like magnets\cite{Rahm},
cylinder-like magnets\cite{Ross} as well as rectangle-like
magnets\cite{Kossut1,Crowell,Bael} are examples of them. Typically,
the structures of arrays of magnets is characterized by Atomic-Force
Microscopy (AFM), the magnetic properties are studied by
Superconducting QUantum Interference Device (SQUID) magnetization
measurements and in order to obtain local microscopic information on
the domain structure of the individual islands, Magnetic-Force
Microscopy (MFM) measurements is performed. Recently we have
presented\cite{Redlinski1,Redlinski2} theoretical analysis of the
absorption spectrum as well as distribution of the local magnetic
field induced in (Ferromagnet)/(DMS) hybrid composed of rectangular
and/or cylindrical island and DMS quantum well structure. We have
shown that depending on the distribution of the magnetic field it is
possible to localize quasi-particle in this type of system.

\begin{figure}[h]
\includegraphics[height=.25\textheight]{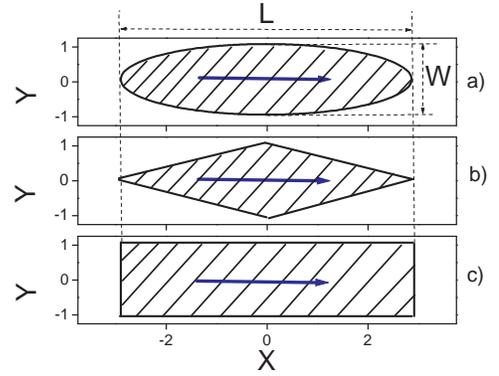}
\caption{Three kinds of flat islands: ellipse, diamond and rectangular shape. Islands
are magnetized uniformly in the x-direction as indicated by arrows. Aspect ratio of the
length $L$ and the width $W$, $L/W=3$, is the same for all three
shapes.}\label{figShapes}
\end{figure}
In the next section I present theoretical assumptions concerning the system as well as
describe procedure of calculation of magnetic field induced by a magnets. Then I
analyze results in more details and draw conclusions.

\section{Theory and Numerical Procedure}
I considered three types of flat islands which volumes $\mathcal{V} = \mathcal{S}\times
H$. Shape $\mathcal{S} \subset \mathbb{R}^2$ is one of ellipse-, diamond- or rectangle
shapes as shown on Fig.~\ref{figShapes} and $H \subset \mathbb{R}$ is height of the
magnet. In my coordinate system, magnet is lying on the XY plane and I assumed that the
aspect ratio between its length $L$ and its width $W$ equals to three, $L/W=3$, and the
aspect ratio between its width $W$ and its height $H$ is assumed to be 10, $W/H=10$.
For example, the following numbers fulfil the assumptions: length $L$=6~$\mu$m, width
$W$=2~$\mu$m and height $H$=0.2~$\mu$m - so magnet is called a micro-magnet. Islands
are magnetized uniformly in the x-direction as indicated on Fig.~\ref{figShapes}.
Z-component of induced magnetic field was calculated at points $\vec{r}=(x, y, -d)$ on
the XY plane at the distance $d=H/20$ below islands.

Magnetic field $\vec{B}(\vec{r})$ is a function of all four lengths:
$L$, $W$, $H$, $d$ and is proportional to the saturation
magnetization $M_s$: $\vec{B}(\vec{r}) \sim M_s\, \vec{f}_0(\vec{r};
L, W, H, d)$. As can be checked by solving magneto-static Maxwell's
equations\cite{Jackson} the following scaling is fulfilled:
$\vec{B}(\vec{r})= M_s \,\lambda\, \vec{f}_0(\vec{r}/\lambda;
L/\lambda, W/\lambda, H/\lambda, d/\lambda)$ which allows us to work
with dimensionless lengths indicated on Fig.~\ref{figShapes} and in
the following presentation.

Magnetic field coming from uniformly magnetized cuboid can be calculated analytically
as shown for example in Ref.~\onlinecite{Engel}. I use this fact and approximate volume
of the whole magnet $\mathcal{V}$ by a sum of volumes of small cuboids $\square_i$,
\mbox{$\mathcal{V} \gtrsim \sum_i \square_i$}, which are totally included in the volume
of island, $\square_i \subset \mathcal{V}$. Then, using the fact that Maxwell equations
are linear, I can approximate total magnetic field $\vec{B}(\vec{r})$ at point
$\vec{r}$ by a sum of partial fields $\vec{B}_{\square_i}(\vec{r})$ produced by each
cuboid $\square_i$:
\begin{equation}\label{sum}
\vec{B}(\vec{r}) \approx  \vec{B}_{appr}(\vec{r})=\sum_i
\vec{B}_{\square_i}(\vec{r}).
\end{equation}
$\vec{B}_{appr}$ that approximate real field $\vec{B}$ (for the rectangular shaped
magnet $\vec{B}_{appr}=\vec{B}$) depends on the division of the volume $\mathcal{V}$. I
used small volumes $a\times a\times H$ to represent each cuboid and assured that
$a<<W$. Typically $W/a=100$ which have given us smooth magnetic field $\vec{B}_{appr}$
after performing summation as required in Eq.~(\ref{sum}). In the further discussion
symbol $\vec{B}$ will be used instead of $\vec{B}_{appr}$ to simplify notation.

\section{Results and Discussion}
In order to compare results obtained for all three shapes I set $L$=6~$\mu$m,
$W$=2~$\mu$m and $H$=0.2~$\mu$m in the case of all types of islands. Magnetic field
$\vec{B}$ was calculated at the same distance $d$=10~nm=0.01~$\mu m$ below each
micro-magnet. Each micro-magnet was magnetized uniformly to the value of the saturation
magnetization of iron $M_s$=1740~$\frac{emu}{cm^3}$. In Tab.~\ref{Ms} I collected
saturation magnetization of few other substances.
\begin{table}
\begin{ruledtabular}
\begin{tabular}{cccc}
Substance & $M_s$ [emu/cm$^3$]& $\mu_0 M_s$ [T] & $\frac{M_s(\text{Substance})}{M_s(\text{Fe})}$ [1]\\
\hline
Fe    &   1740 (1707)  &  2.19 (2.15)   &  1.00 \\
Co    &   1446 (1400)   &  1.82 (1.76)  &  0.83 \\
Ni    &   510 (485)    &   0.64 (0.61)  &  0.29 \\
Dy    &   2920 (-)    &    3.67 (-)     &  1.68 \\
MnAs  &   870 (670)    &   1.1 (0.84)   &  0.50\\
MnSb  &   - (710)   & - (0.89)          &  0.41
\end{tabular}
\end{ruledtabular}
\caption{Saturation magnetization at T=0~K of different substances
in CGS units and SI units taken from Ref.~\onlinecite{Kittel}. Room
temperature values are shown in brackets. Results presented in the
report scale easly with $M_s$, e.g, when we use $M_s$ of Nickel
instead of $M_s$ of Fe then induced field is almost four times
smaller.}\label{Ms}
\end{table}
Results scales simply with $M_s$ and it is straightforward to obtain distribution of
the magnetic field of other substances by multiplying results by the factor
$M_s$(Substance)/$M_s$(Fe) which I show in the last column of Tab.~\ref{Ms}. For
example, when considering MnAs island instead of Fe island, the results have to be
multiplied by a factor of $M_s(\text{MnAs})/M_s(\text{Fe})=870/1740 = 0.5$.

\begin{figure}[h]
\includegraphics[height=.25\textheight]{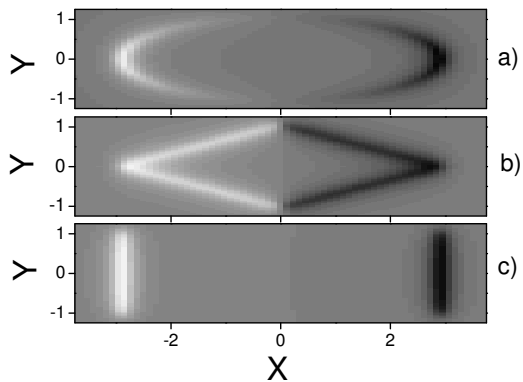}
\caption{Contour plot of the $B_z$ of three kinds of shapes: ellipse, diamond and
rectangular in the XY plane at distance $d$=10~nm below iron islands. I used gray
scale: black (white) region denotes negative (positive) values of $B_z$. Figure
reflects the distribution of the magnetic charge on the surface of
islands.}\label{fig2D}
\end{figure}

Fig.~\ref{fig2D} shows contour plot of the z-component of the
magnetic field, $B_z$, of ellipse shape micro-magnet (panel a), of
diamond shape micro-magnet (panel b) and of rectangle shape magnet
(panel c). White (black) color indicates region with positive
(negative) value of $B_z$ and gray color indicates $B_z=0$.
As seen on Fig.~\ref{fig2D} distribution of magnetic field on the XY plane is highly
position-dependant and shape-dependant. Large fields are induced in proximity of
magnetic poles below and above them. In the further discussion I will use
magneto-static language and use notion of the density of magnetic charge $m(\vec{r})$.
Total magnetic charge is the same for all shapes of islands because I have chosen
identical $M_s$ for all types micro-magnets. Magnetization
$\vec{M}(\vec{r})=M_s(\vec{r})\vec{e}_x$, and density of magnetic charge $m(\vec{r})
\equiv \vec{M}(\vec{r}) \cdot d\vec{S}(\vec{r})$ where $d\vec{S}(\vec{r})$ is a unit
vector normal to the surface at point $\vec{r}$. It means that in the case of diamond
and rectangle shapes $\vec{m}$ is a constant on the surface of micro-magnets because
normal to its surface does not depend on the position. In the case of diamond-shape
magnet $\vec{m}$ is $\sqrt{10}$ times smaller than $\vec{m}$ of rectangle-shape magnet
because surface area is $\sqrt{10}$ time larger in the case of diamond-shape magnet.
From this fact we expect that magnitude of $B_z$ is about $\sqrt{10}\approx 3.2$
smaller in the case of diamond shape then of rectangle shape. As we shall see later
three dimensional plots confirm this fact. In the case of ellipse-shape magnet normal
to the surface changes with position on the surface so does $m(\vec{r})$.
Fig.~\ref{fig2D} reflects the distribution of magnetic charge.

To gain more knowladge about distribution of the field now I present three dimensional
plots of $B_z$ in the region of its large value.
\begin{figure}[h]
\includegraphics[height=.25\textheight]{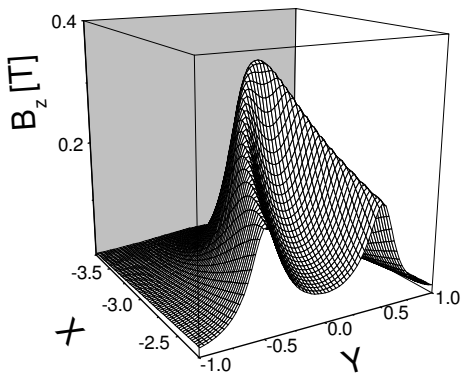}
\caption{Distribution of the $B_z$ around magnetic pole of
ellipse-like micromagnet ($L$=6~$\mu$m, $W$=2~$\mu$m,
$H$=0.2~$\mu$m) in the XY plane at distance $d$=10~nm below the
magnet. Induced field is localize at point (0, -3) on XY
plane.}\label{figOval}
\end{figure}
Figures~\ref{figOval}, \ref{figDiam} and \ref{figRect} show z-component of the magnetic
field, $B_z$, that is produced by the micro-magnets of ellipse, diamond and rectangle
shape, respectively. Induced fields can be directly compared because all three scales
$(X,Y,B_z)$ are the same on the figures. In all three cases, as before, magnetic field
is calculated at the distance $d$=10~nm below micro-magnets and, as before,
$L$=6~$\mu$m, $W$=2~$\mu$m, $H$=0.2~$\mu$m. I plotted only region below left pole in
its close proximity. On the right pole $\vec{B}_z$ has opposite sign. We see again that
in the case of ellipse magnet, Fig.~\ref{figOval}, induced field is highly localized
around point (0, -3). Magnitude of $B_z$ is almost 0.4~T and is almost the same as in
the case of rectangular island, Fig.~\ref{figRect}.
\begin{figure}[h]
\includegraphics[height=.25\textheight]{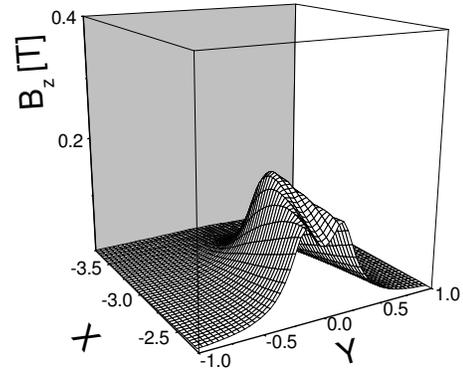}
\caption{Distribution of the $B_z$ around magnetic pole of
diamond-like micromagnet ($L$=6~$\mu$m, $W$=2~$\mu$m,
$H$=0.2~$\mu$m) in the XY plane at distance $d$=10~nm below island.
The largest value of $B_z$ is found at point (0,
-3).}\label{figDiam}
\end{figure}
In the case of diamond-like magnet, Fig.~\ref{figDiam}, magnetic
field is localized around point (0, -3), too, but the amplitude is
much lower then in the case of ellipse magnet. There exist large
component of the magnetic field below edges of the whole magnet.
\begin{figure}[h]
\includegraphics[height=.25\textheight]{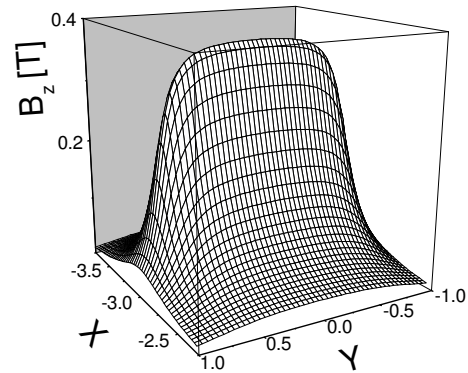}
\caption{Distribution of the $B_z$ around magnetic pole of
rectangle-like micromagnet ($L$=6~$\mu$m, $W$=2~$\mu$m,
$H$=0.2~$\mu$m) in XY plane at distance $d$=10~nm below the magnet.
Large $B_z$ component exists below and above whole left and right
edge of island.}\label{figRect}
\end{figure}
In the case of rectangular magnet large $B_z$ component exists below whole left and
right edge of the island. As was shown in Ref.~\onlinecite{Redlinski2} such elongation
of the local field can produce one-dimensional traps for quasi-particles. Presented
analysis and figures allow us to conclude that in the case of ellipse sample
zero-dimensional spin-traps can be realized in the (Ferromagnet)/(DMS) hybrid.

\section{Conclusions}
I analyzed distribution of the magnetic field induced by micro- and nano-magnets of
three different shapes: ellipse, diamond and rectangle. As a main results I have found
that the shape of islands has large impact on the distribution of induced field. My
calculations show that ellipse-shaped island produces much more localized fields in
proximity of its magnetic poles in contrast to rectangular-shaped island. Additionally,
magnitude of this field is large and almost the same as a magnitude of the field
induced by rectangular shape. I expect that (Ferromagnet)/(DMS) hybrid composed of
ellipse micro-magnet and DMS quantum well structure will produce zero-dimensional traps
for quasi-particle.

\section{Acknowledgement}
I would like to thank Dr K. Lebecki for useful discussions
concerning this topic as well closely related problems.

This material is based in part upon work supported by the NSF
DMR02-10519. Any opinions, findings, and conclusions or
recommendations expressed in this material do not necessarily
reflect the views of the NSF.


\end{document}